\documentclass{article}

\usepackage{arxiv}

\usepackage[utf8]{inputenc} 
\usepackage[T1]{fontenc}    
\usepackage{hyperref}       
\hypersetup{
    colorlinks=true,
    linkcolor=blue,
    filecolor=magenta,      
    urlcolor=cyan,
    pdftitle={CaNS-Fizzy},
    pdfsubject={physics.flu-dyn},
    pdfauthor={G.~Lupo, P.~Costa, P.~Wellens},
    pdfpagemode=FullScreen,
    citecolor=black
    }
\usepackage{url}            
\usepackage{booktabs}       
\usepackage{amsfonts}       
\usepackage{nicefrac}       
\usepackage{microtype}      
\usepackage{graphicx}
\usepackage{natbib}
\usepackage{doi}

\title{CaNS-Fizzy: A GPU-accelerated finite difference solver for turbulent two-phase flows}

\author{ \href{https://orcid.org/0000-0002-1095-118X}{\includegraphics[scale=0.06]{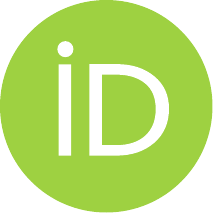}\hspace{1mm}Giandomenico~Lupo}\thanks{Corresponding author} \\
	Delft University of Technology\\
    Department of Process \& Energy\\
    Delft, The Netherlands\\
	\texttt{G.Lupo@tudelft.nl} \\
	\And
	\href{https://orcid.org/0000-0003-0009-9058}{\includegraphics[scale=0.06]{orcid.pdf}\hspace{1mm}Peter~Wellens} \\
	Delft University of Technology\\
    Department of Marine \& Transport Technology\\
    Delft, The Netherlands\\
	\texttt{P.R.Wellens@tudelft.nl} \\
	\And
	\href{https://orcid.org/0000-0001-7010-1040}{\includegraphics[scale=0.06]{orcid.pdf}\hspace{1mm}Pedro~Costa} \\
	Delft University of Technology\\
    Department of Process \& Energy\\
    Delft, The Netherlands\\
	\texttt{P.SimoesCosta@tudelft.nl} \\
}

\date{}

\renewcommand{\undertitle}{}
\renewcommand{\shorttitle}{CaNS-Fizzy}

\begin{document}
\maketitle

\begin{abstract}
	
\href{https://github.com/CaNS-World/CaNS-Fizzy}{CaNS-Fizzy} -- Fizzy for short -- is a GPU-accelerated numerical solver for massively-parallel Direct Numerical Simulations (DNS) of incompressible two-phase flows. A DNS enables direct access to all flow quantities, resolved in time and space at all relevant continuum scales. The resulting numerical experiments provide complete data sets for the analysis of the detailed mechanisms underlying the flow, particularly the interaction between the chaotic and multi-scale dynamics of turbulence and the interface movement and deformation. The insights gained can guide the design and operation of various applications, such as boiling heat transfer, liquid-liquid extraction, gas-liquid reactors, absorption and stripping columns, distillation columns, liquid combustion appliances, in all of which the rate of heat and mass transfer between phases is proportional to the interfacial area. Fizzy's two-phase capabilities were implemented using the efficient, GPU-accelerated Navier-Stokes solver CaNS \citep{Costa:2018} as base.

\end{abstract}

\section{Statement of need}

\href{https://github.com/CaNS-World/CaNS-Fizzy}{Fizzy} is suited for large-scale direct numerical simulations of canonical incompressible two-phase flows, from simple laminar cases to bubble/droplet-laden turbulent suspensions. These flows may be computationally expensive due to the stringent resolution requirements imposed by the direct solution of immersed interfaces dispersed throughout the domain. This demands efficient use of the capabilities of modern computing systems; Fizzy has been developed to include key desirable features that enable this objective: a one-fluid formulation of the two-phase flow governing equations, the use of a fast direct solver for the pressure Poisson equation, and an efficient distributed GPU porting with an interface capturing strategy that is suitable for GPU acceleration.
In addition to the momentum transfer and interface capturing, the code has the capability to solve heat transfer in both fluid phases, and thermal convection based on the Oberbeck-Boussinesq approximation.
Finally, the code has been extensively validated with several benchmark cases that demonstrate the different features of the solver, which are incorporated in the continuous integration workflows of the repository.

\section{Mathematical model}

A one-fluid formulation of the two-phase flow is employed, solving a single set of governing equations for both phases in the whole domain. The incompressible Navier-Stokes equation, the heat transport equation and the Accurate Conservative Diffuse Interface (ACDI) transport equation are evolved in time to compute the velocity and pressure, temperature and phase indicator fields respectively. The latter identifies the regions of the domain occupied by either phase: it is continuous and smooth over the whole domain, and the interface between phases is diffuse. The thermophysical and transport properties (density, viscosity, thermal conductivity, specific heat capacity) are linearly mapped over the phase indicator field, and thus also continuous and smooth. The surface tension at the interface is included as a Continuous Surface Force (CSF) \citep{Brackbill:1992} in the Navier-Stokes equation.
See \citet{Costa:2018} and \citet{Jain:2022} for more details.

\section{Methods and Implementation strategy}

The governing equations are spatially discretized with a second-order finite difference scheme on a 3D Cartesian grid; a staggered grid arrangement is used for the velocity field, while all other quantities are stored at the cell centers; time integration is based on a low-storage three-step Runge-Kutta scheme. The incompressible Navier-Stokes equation is solved with a pressure correction scheme to enforce mass conservation, which yields a variable coefficient Poisson equation for the pressure correction: a splitting technique adapted from \citet{Dong:2012} and \citet{Dodd:2014} transforms this equation into a constant coefficient Poisson equation \citep{Frantzis:2019}, enabling the use of the fast direct FFT solver of the CaNS code. Fizzy also allows for solving the conventional variable-coefficients problem using a geometric multigrid method through the \href{https://github.com/hypre-space/hypre}{Hypre} library.
The diffuse interface representation of the phase interface allows for continuous and smooth mapping of the physical fields across the interface, and it requires no explicit interface reconstruction thanks to the interface regularization flux, keeping the computational load constant regardless of the local interface topology and thus making the algorithm particularly suited for parallelization on GPU architecture. The momentum equation includes the flux associated with the diffuse interface regularization, which allows for a mass--momentum consistent discretization and enables stability at high density contrasts between phases.

The code is written in modern Fortran, and is parallelized using MPI and OpenACC directives for GPU kernel offloading and host/device data transfer. As in CaNS, Fizzy leverages \href{https://github.com/NVIDIA/cuDecomp}{cuDecomp} \citep{Romero:2022} for distributed memory calculations in pencil domain decompositions, and \href{https://docs.nvidia.com/cuda/cufft/}{cuFFT} for computing Fourier transforms. On CPUs, the code uses \href{https://github.com/2decomp-fft/2decomp-fft}{2DECOMP\&FFT} \citep{Rolfo:2023} and \href{https://www.fftw.org/}{FFTW} to perform the same operations.

Users can design and run a simulation by specifying the physical and computational parameters in a simple Fortran namelist input file. The code uses a modular, procedural design which makes extensions with different numerical methods or physical phenomena easy to develop. In the short term, we aim to allow for different interface tracking algorithms (e.g., based on the volume-of-fluid method), along with alternative schemes for spatial and temporal discretization.

Finally, the code was designed so that important new computational features in the parent solver CaNS (e.g. porting efforts to other architectures) are easily propagated to Fizzy.

\section{Examples}

\begin{figure}[h!]
	\centering
    \includegraphics[width=0.75\textwidth]{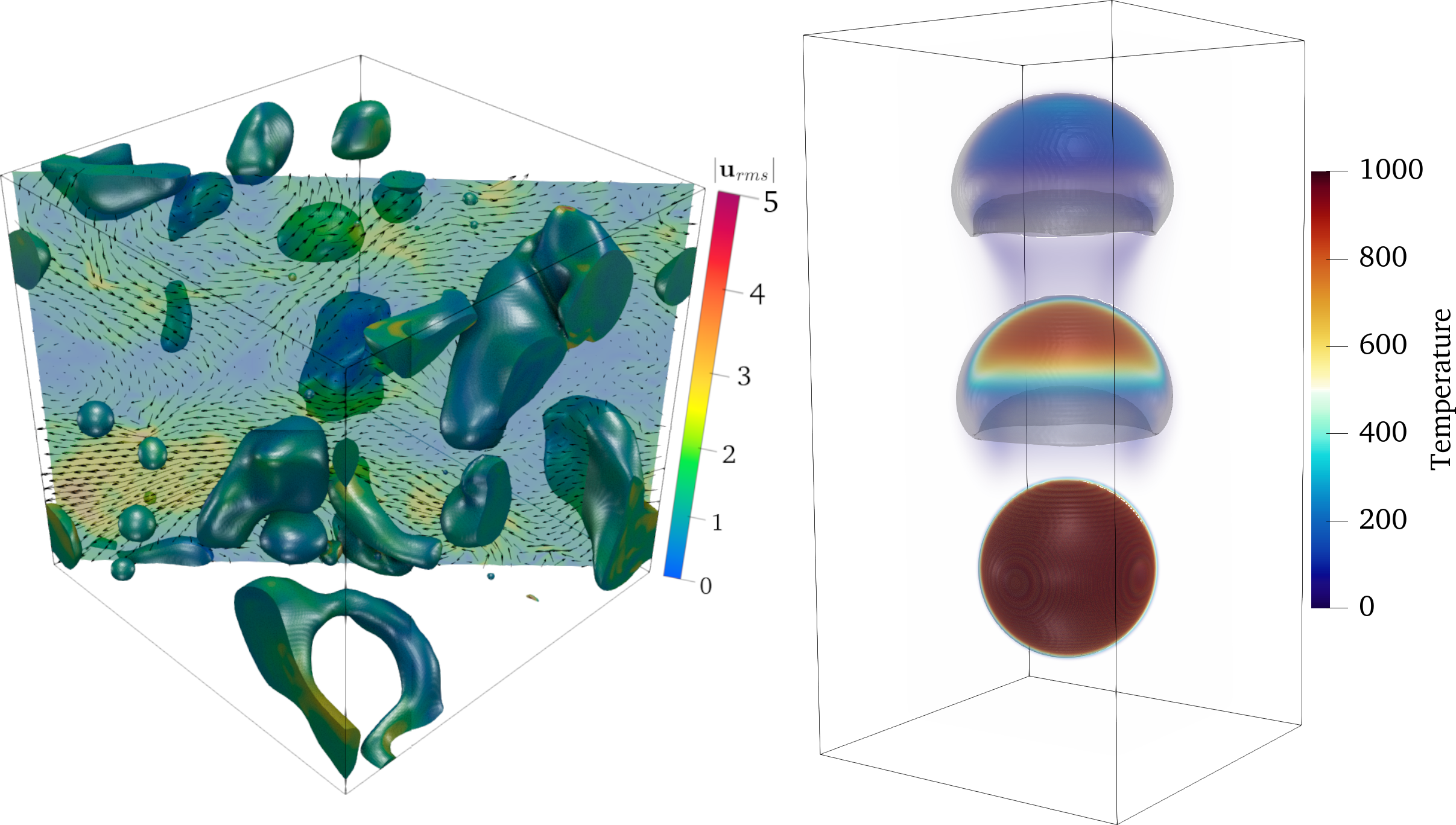}
	\caption{(Left) Simulation of a liquid-liquid emulsion in a three-periodic domain with sustained homogeneous isotropic turbulence. The colour represents the velocity magnitude. A diagonal plane with in-plane velocity vectors is shown. (Right) Three successive snapshots of a hot gas bubble rising in a cold liquid. A coloured volume rendering of the temperature field is shown both in the gas bubble and in the liquid wake.}
	\label{fig:examples}
\end{figure}

Figure~\ref{fig:examples} illustrates examples of two-phase flows simulated with this solver. The left panel shows how forced homogeneous isotropic turbulence breaks and deforms the interface of a liquid-liquid emulsion in a triperiodic domain; the right panel shows a hot gas bubble rising in a cold liquid, attaining the typical skirt shape while cooling down and simultaneously heating up the surrounding liquid in its wake.

\section{Computational performance}

\begin{figure}[h!]
	\centering
    \includegraphics[width=0.45\textwidth]{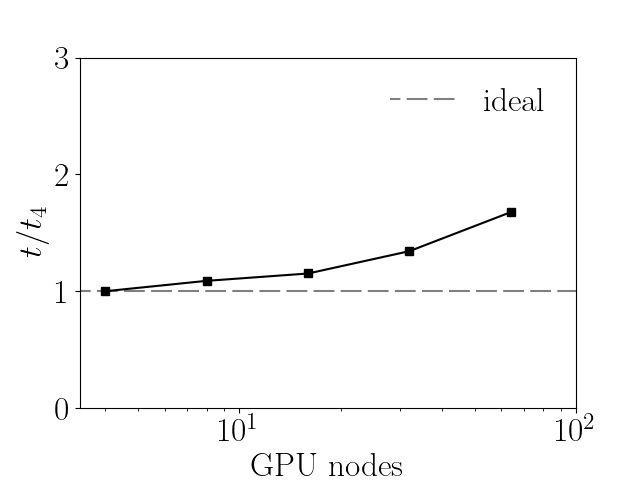}
	\caption{Weak scaling performance on GPU nodes at full memory. The vertical axis shows the wall-clock time normalized by the four-node case.}
	\label{fig:performance}
\end{figure}

Fizzy is tailored for large-scale simulations that exploit the computational capacity of modern GPU clusters with full GPU occupancy. The most relevant metric of the parallel efficiency for such scenario is a weak scaling test that determines the penalty in increased wall-clock time occurring when the problem size is increased alongside the computational resources. The liquid-liquid emulsion in homogeneous isotropic turbulence case of Figure~\ref{fig:examples} (Left) has been used for this test: the size of the computational domain has been extended in one direction linearly with the number of GPU nodes employed. The test has been carried out on the GPU partition of the supercomputer Leonardo from Cineca, Italy; each computing node is equipped with four NVIDIA A100 SXM6 64GB GPUs, and is able to fit at full memory a $1024^3$ ($\sim$ 1 billion grid cells) computational box. Figure~\ref{fig:performance} shows the performance penalty as the problem domain size (i.e. the number of spatial degrees of freedom) is increased from occupying 4 nodes (16 GPUs) to 64 nodes (256 GPUs): the 16 times larger computation takes only about 1.7 times longer than the original 4-node computation.
The key contributor to the parallel performance is the interface capturing approach used in the ACDI method, which prevents thread divergence in GPU kernels, as the computational load is independent of the local interface morphology due to the lack of explicit interface reconstruction. Indeed, very little sensitivity of the wall-clock time per iteration to the amount of interface area is observed, even for unsteady evolution of the interface with drastic topology changes during break-up events.

\section*{Acknowledgements}

We would like to thank Jordi Poblador-Ibanez, Suhas Jain, Naoki Hori, and Bergmann Óli Aðalsteinsson for insightful discussions that led to practical improvements of the numerical method.
This work was partially supported by a Cohesion Grant from the Mechanical Engineering Faculty at TU Delft awarded to Pedro Costa and Peter Wellens.

\bibliographystyle{unsrtnat}
\bibliography{references}

\begin{thebibliography}{8}
\providecommand{\natexlab}[1]{#1}
\providecommand{\url}[1]{\texttt{#1}}
\expandafter\ifx\csname urlstyle\endcsname\relax
  \providecommand{\doi}[1]{doi: #1}\else
  \providecommand{\doi}{doi: \begingroup \urlstyle{rm}\Url}\fi

\bibitem[Costa(2018)]{Costa:2018}
P.~Costa.
\newblock A {FFT}-based finite-difference solver for massively-parallel direct
  numerical simulations of turbulent flows.
\newblock \emph{Comput. Math. Appl.}, 76\penalty0 (8):\penalty0 1853--1862,
  2018.
\newblock \doi{10.1016/j.camwa.2018.07.034}.

\bibitem[Brackbill et~al.(1992)Brackbill, Kothe, and Zemach]{Brackbill:1992}
J.~U. Brackbill, D.~B. Kothe, and C.~Zemach.
\newblock A continuum method for modeling surface tension.
\newblock \emph{J. Comput. Phys.}, 100\penalty0 (2):\penalty0 335--354, 1992.
\newblock \doi{10.1016/0021-9991(92)90240-Y}.

\bibitem[Jain(2022)]{Jain:2022}
S.~S. Jain.
\newblock Accurate conservative phase-field method for simulation of two-phase
  flows.
\newblock \emph{J. Comput. Phys.}, 469:\penalty0 111529, 2022.
\newblock \doi{10.1016/j.jcp.2022.111529}.

\bibitem[Dong and Shen(2012)]{Dong:2012}
S.~Dong and J.~Shen.
\newblock A time-stepping scheme involving constant coefficient matrices for
  phase-field simulations of two-phase incompressible flows with large density
  ratios.
\newblock \emph{J. Comput. Phys.}, 231\penalty0 (17):\penalty0 5788--5804,
  2012.
\newblock \doi{10.1016/j.jcp.2012.04.041}.

\bibitem[Dodd and Ferrante(2014)]{Dodd:2014}
M.~S. Dodd and A.~Ferrante.
\newblock A fast pressure-correction method for incompressible two-fluid flows.
\newblock \emph{J. Comput. Phys.}, 273:\penalty0 416--434, 2014.
\newblock \doi{10.1016/j.jcp.2014.05.024}.

\bibitem[Frantzis and Grigoriadis(2019)]{Frantzis:2019}
C.~Frantzis and D.~G.~E. Grigoriadis.
\newblock An efficient method for two-fluid incompressible flows appropriate
  for the immersed boundary method.
\newblock \emph{J. Comput. Phys.}, 376:\penalty0 28--53, 2019.
\newblock \doi{10.1016/j.jcp.2018.09.035}.

\bibitem[Romero et~al.(2022)Romero, Costa, and Fatica]{Romero:2022}
J.~Romero, P.~Costa, and M.~Fatica.
\newblock Distributed-memory simulations of turbulent flows on modern {GPU}
  systems using an adaptive pencil decomposition library.
\newblock In \emph{Proceedings of the Platform for Advanced Scientific
  Computing Conference}, PASC '22, New York, NY, USA, 2022. Association for
  Computing Machinery.
\newblock \doi{10.1145/3539781.3539797}.

\bibitem[Rolfo et~al.(2023)Rolfo, Flageul, Bartholomew, Spiga, and
  Laizet]{Rolfo:2023}
S.~Rolfo, C.~Flageul, P.~Bartholomew, F.~Spiga, and S.~Laizet.
\newblock The {2DECOMP\&FFT} library: an update with new {CPU/GPU}
  capabilities.
\newblock \emph{J. Open Source Softw.}, 8\penalty0 (91):\penalty0 5813, 2023.
\newblock \doi{10.21105/joss.05813}.

\end{thebibliography}

\end{document}